# エッジ画像分析及び ERP を用いた万引防止サービスの検討

山登庸次[†]　福本佳史[†]　熊崎宏樹[†]

† NTT ソフトウェアイノベーションセンタ
東京都武蔵野市緑町 3-9-11
E-mail: †{yamato.yoji,fukumoto.yoshifumi,kumzaki.hiroki}@lab.ntt.co.jp

**あらまし** 本稿では，監視カメラ映像分析と ERP を用いて，万引行為を検出し，万引を防止するための，SaaS サービスを提案する．日本での万引被害は 5000 億円近くで，高価な対策を導入できない小規模店舗が年間 1000 以上廃業に追い込まれている．本稿では，近年発展している，クラウド技術，データ分析技術を用いて，小規模店舗でも導入可能な，監視カメラ映像分析結果と ERP データ突合を用いた，万引行為を検出するサービスを提案する．更に，機械学習フレームワーク Jubatus を用いて，監視カメラ映像のストリーム処理の実現性を確認する．
**キーワード** 万引防止，Jubatus，クラウド，画像分析，ERP

## Study of shoplifting prevention using image analysis and ERP check

Yoji YAMATO[†], Yoshifumi FUKUMOTO[†], and Hiroki KUMAZAKI[†]

† Software Innovation Center, NTT Corporation
3-9-11, Midori-cho, Musashino-shi, Tokyo 1808585 Japan
E-mail: †{yamato.yoji,fukumoto.yoshifumi,kumzaki.hiroki}@lab.ntt.co.jp

**Abstract** In this paper, we propose a SaaS service which prevents shoplifting using image analysis and ERP. In Japan, total damage of shoplifting reaches 450 billion yen and more than 1000 small shops gave up their businesses because of shoplifting. Based on recent cloud technology and data analysis technology, we propose a shoplifting prevention service with image analysis of security camera and ERP data check for small shops. We evaluated stream analysis of security camera movie using online machine learining framework Jubatus.

## 1. はじめに

近年，クラウド技術，データ分析技術が進展している．SaaS (Software as a Service) 型クラウドサービスでは，クラウド事業者は，ERP (Enterprise Resource Planning) 等の業務系アプリケーションや，メール，Web 等の情報系アプリケーションを，ネットワーク経由で高可用に提供しており，ユーザはクラウド事業者と契約することで，自らインフラは持たなくてもこれらのアプリケーションをサービスとして安定的に利用できる．（クラウド事業の開発例として例えば [1]，クラウド高可用技術の例として例えば [2]）

また，クラウドの計算リソースを用いて，データを分析する技術も普及している．複数顧客の購買履歴から顧客に商品をリコメンドするリコメンドサービスや，店舗内の監視カメラに映る人の動きから人の動線を分析するサービス等が，Hadoop (HDFS [3] 等複数機能で構成される) や Apache Spark [4] 等のビッグデータ分析技術を用いて提供されている．

一方，実店舗での商品販売を見ると，小売店での万引被害は，日本では年間 4500 億円以上と報告されており [5]，ドラッグストア，スーパーマーケット，書店等での被害は各 100 億円以上に及んでいる．万引を防止するために，店舗側の対策としては，店舗内を監視する人を増やす，店舗を監視するカメラを人の目で常に見て問題が無いか確認する，商品にタグを添付し販売していない商品の持出を店舗入り口のゲートでアラートする等の対策がされている．しかしこれらの対策は，監視する人の人件費や，アラートゲート設置等の初期コストが高くつくため，小規模の店舗では実施できないのが現状である．例えば，書店は小規模店舗が多いが，小規模書店ではこれらの対策は打てず，万引被害が原因で年間 1200 店舗が廃業に追い込まれている．

これらの背景を踏まえ，本稿では，近年発展している，クラウド技術，データ分析技術を用いて，小規模店舗でも導入可能な，低コストな万引防止サービスの実現を目的とする．本稿提案技術は，店舗側に配置された小型コンピュータ上の機械学習フレームワーク Jubatus [6] が，監視カメラ映像を分析し異常を

検出してクラウドに通知し，クラウド上の万引防止アプリケーションが，ERP 等の商品数データと突合し，万引の疑いが高い場合に，店舗スタッフにメール等で通知する．

## 2. 従来の技術と課題

追加人件費が不要で，万引を防止するためのシステムとして，サブローくん A [7] がある．サブローくん A は，監視カメラ映像から，客の動作が事前に定義した 50 種類以上の不審行動項目に該当した場合に，検出し，店員に通知を行う．店員は，声掛けなどをして，注意喚起をすることで，万引を未然に防ぐことを目指している．しかし，サブローくん A では，映像解析用のソフトや PC を店舗に購入しておく必要があり初期コストがかかること，事前に定義した不審行動項目以外の新たな万引行為を防げないこと，不審動作の検出精度は 100%でないため万引でないのに声掛けする必要が頻発し，店舗の運用上厳しい場合があること，の課題がある．

商品が，特定エリアから消失したこと，および復帰したことを，カメラ映像から検知して万引を検出する，パナソニック株式会社の技術 [8] がある．この技術は，カメラによる画像情報に基づいて，対象物配置エリアから対象物が消失したこと及び商品配置エリアに対象物が復帰したことを検知する対象物検知部と，この対象物検知部の検知結果に基づいて対象物の交換を検知する交換検知部と，この交換検知部の検知結果に基づいて分析結果となる出力情報を生成する出力情報生成部とを備える．しかし，この技術は，対象物の消失，復帰を判定することが主目的であるため，万引行為か購買行為か判定ができないため，長時間が経って万引行為であったことを確認する使い方となるため，万引行為が起きた際に捕まえたり，注意喚起することは難しい．

以上のように，従来技術は，事前に定義したルールに基づく万引しか検出できない場合があること，カメラ映像分析だけでは精度が十分でないため声掛けなどの実運用が難しい場合があること，の 2 つの課題がある．そこで，本稿では，小規模な店舗でも万引防止を行えるようにするため，事前ルールに無い万引行為も含めて高い精度で検出し，店員に万引行為を通知する，SaaS サービスの提供を目的とする．

## 3. 監視カメラ画像分析と ERP を用いた万引防止サービスの提案

### 3.1 課題を解決するためのアイデア

前節で記載した事前ルールに無い検出が困難，検出精度が不十分の課題に対応するため，以下のアイデアで解決を図る．

事前ルールに無い万引行為にも対応するため，機械学習技術により，運用データに基づき不自然な行動であれば検出可能にする．精度高い検出とするため，カメラ画像だけでなくクラウドで管理される商品 DB と突合することで，判定の確度を高める．これらのアイデアに基づいた，システム構成と処理ステップを説明する．

### 3.2 提案技術の構成図

図 1 に，提案技術のシステム構成図を示す．システムは，商

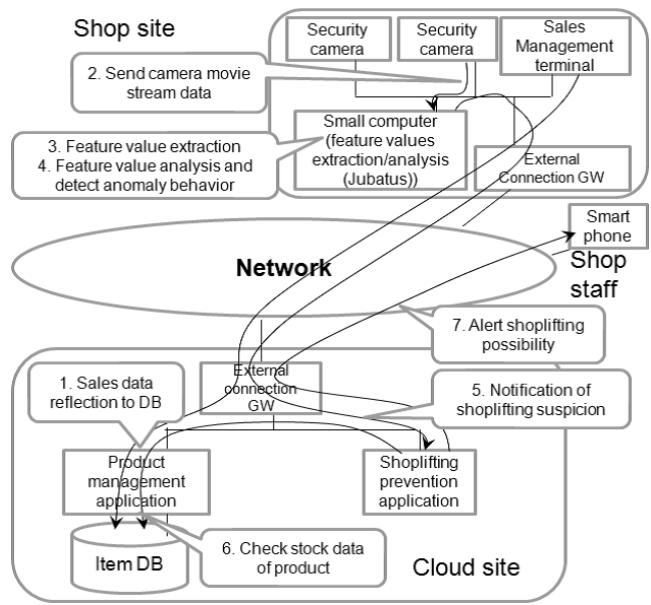

図 1　システム構成図と処理ステップ

品データ等を管理するクラウド側と，監視カメラが配置される各店舗（エッジ）側が，ネットワークにより接続されている．ネットワークは Internet や必要に応じてセキュアな VPN を用いる．

店舗側には，監視カメラと小型コンピュータと売上管理端末と外部接続 GW が配置され，ネットワークに外部接続 GW を介して接続される．監視カメラは，店舗に配置されるネットワークカメラであり，店舗内の映像をストリームデータで小型コンピュータに送る．小型コンピュータは機械学習フレームワーク Jubatus が搭載され，カメラ映像を分析して万引疑い行為を検出しクラウド側に通知する．売上管理端末は，商品販売を管理する端末で，販売データは，SaaS である商品アプリケーションに通知され，商品 DB に反映される．また，店員は，万引疑い通知用の携帯端末を保持する．

クラウド側には，万引防止アプリケーションと，商品管理アプリケーションと商品 DB と，外部接続 GW が配置され，ネットワークに外部接続 GW を介して接続される．万引防止アプリケーションは，万引疑い通知に応じて，商品 DB の商品在庫と実際の商品数に不整合が無いかを照合することで，万引疑い判定の精度を高め，高い確率で万引疑いの場合に，店員携帯端末に通知する．商品管理アプリケーションは，商品の売上等の情報通知に伴い，随時商品 DB に反映する．

### 3.3 提案技術の処理ステップ

図 1 を参照しながら，提案技術の処理ステップを説明する．

ステップ 1：万引の検出とは独立して，売上管理端末は，ネットワークを介して，売上情報をクラウドサイト側の商品管理アプリケーションに通知する．

ステップ 1 は商品管理アプリケーションに繋がれば良いため，POS 端末やクライアント PC 等の既存の売上管理端末が利用できる．商品管理アプリケーションは，ERP 等の業務アプリケーションソフトが SaaS として提供されているもので，

クラウド側で，売上情報や商品在庫情報等が，商品 DB に反映される．商品在庫等の情報は，店舗での販売に応じて，随時商品 DB に反映される．SaaS アプリケーションによっては，この反映が定時のバッチ処理の場合もある．しかし，商品 DB のデータは，万引疑いがある場合の照合にも用いるため，リアルタイムに反映されることが必要である．リアルタイムに商品データを反映するため，商品 DB は高速の処理が可能なシステム，例えば，VoltDB [9] のようなインメモリ DB を用いたシステムを利用する．なお，商品管理アプリケーションとしては，ADempiere [10] 等のオープンソースの ERP が利用できる．

ステップ 2：監視カメラ映像のストリームデータは，店舗内ネットワークを介して，店舗の小型コンピュータに送られる．

小型コンピュータは，例えば Rasbpberry Pi のように一定の計算能力，記憶能力，通信能力を持つコンピュータで，千円前後で購入が出来る．

ステップ 3：小型コンピュータは，映像ストリームデータを，個々の画像データとして切り出し，画像データから特徴値を抽出する．

特徴値とは，画像の色，輝度，輪郭，固有値，物体の形状，数，等の特徴となるデータである．映像データからの特徴値抽出は，dlib [11], OpenCV [12] 等のライブラリを用いることで抽出できる．

ステップ 4：小型コンピュータは，抽出された特徴値から，万引に該当するような行動を検出する．ストリームで随時データが生成される特徴値を分析する機械学習フレームワークとして Jubatus を用いる．

Jubatus は，Apache Storm [13] のようにストリームで分析処理するフレームワークである．Jubatus は，サブローくん A のように事前に定義されたルールに従って，万引に該当する行動を検出するだけでなく，オンラインで新たなルールを学習することもできる．具体的には，通常のカメラ映像との差分が大きい場合に，差分を異常として検出することで，事前定義ルールにない場合もアノマリー判定ができる．アノマリー検出には LOF (Local Outlier Factor) [14] 等のアルゴリズムが用いられる．例えば，バッグに商品を入れる行動は事前ルールとして定義されているが，商品を洋服の中に入れる行動が事前ルールに定義されていない場合に，サブローくん A では検出できないが，Jubatus では通常のカメラ映像との差分からアノマリーとして検出できる可能性がある．

ステップ 5：アノマリースコアが高い等，万引疑いが Jubatus で検知された場合は，その情報が，クラウド側の万引防止アプリケーションに通知される．通知には，http や MQTT (Message Queueing Telemetry Transport) [15] 等を用いれば良い．

ステップ 6：万引防止アプリケーションでは，画像データの分析だけでは，万引の確度が十分でないため，万引疑いの確度を上げるため，商品管理アプリケーションを介して，商品 DB の商品在庫数との照合を行う．

ステップ 1 で商品在庫数は，商品販売と連動して随時更新されているため，本来，商品が何個あるべきかという情報がリアルタイムに更新されている．一方，店舗の商品棚には，万引されている場合，実際の商品の数が商品在庫数と不整合となる．商品棚の実際の商品の数は，監視カメラ画像から，万引行為と同様に，ステップ 3, 4 の分析を行い，数を画像認識することで取得可能である．監視カメラ画像から商品数の判定には，例えば，パナソニック株式会社技術 [8] などの既存技術もある．あるいは，商品棚に数や重量を計測するセンサを設置し，センサにより数を算出しても良い．

万引防止アプリケーションは，商品在庫数と商品棚の実数が不整合があることを確認する．

ステップ 7：万引防止アプリケーションは，画像分析と商品在庫数照合の両方で万引疑いがある場合に，万引疑いが高いとして，店員の携帯端末に，万引疑い客の画像データ及び万引可能性がある商品情報を通知する．

店員は携帯端末の通知を見て，万引疑い客に声掛け，質問などをして，万引を防止する．

## 4. Jubatus により監視カメラ画像分析の簡易評価

監視カメラ映像を Jubatus でストリーム処理で分析した際にどの程度の精度が出るかの確認のため，簡易評価を行った．物を盗むという挙動の推定のため，まずは，人が向いている方向等の状況を，監視カメラの切り出し画像から Jubatus により判定する Python クライアント，およびカメラ映像の切り出し画像から特徴抽出を行う Jubatus プラグインを実装した．

特徴抽出のため，顔から，目・鼻・口・輪郭の座標を 68 点抽出する dlib ライブラリを用いた．dlib は，画像処理の機能を持つライブラリで顔検出等に実績が多い．二次元上で得られる座標点 68 点から，x 軸 y 軸で分け，合計 136 個の特徴値とし，136 次元ベクトルとする．この特徴値に対し，顔画像の位置を差し引きサイズで割ることで，顔画像上での相対座標へと正規化を行う．正規化されたデータを，Jubatus の classifier 機能を利用し，線形分類器 (Linear Classifier) と kNN [16] 分類器 (k Nearest Neighbor Classifier) で分類し，精度の良い方を判定結果として利用する．図 3 に簡易評価の概要を示す．

実際の人間が映っている画像 1103 件から，「前を向いている」「目を閉じている」「下を向いている」「横を向いている」の判定を行い，k=10 での k-交差検証を行ったところ，正解率は 72%であった．今回の検証は簡易のため未チューニングの部分が多い．この簡易検証から，カメラ映像からのストリームでの画像分析で姿勢等をある程度の正解率で判定可能な点が分かる．更に，監視カメラでの分析だけでは正解率にも限界があることや，店舗ごとのチューニングが必要な点もあることから，ERP 等の商品 DB の商品データとの突合により，より万引検出の精度を上げる必要がある点も分かる．

なお，Jubatus で判定する際の元となる学習モデル等は，クラウドの万引防止アプリケーションから各店舗の小型コンピュータに配信することを想定している．学習モデルを含め設定等の更新時は，例えば [17] のようなクラウドの一括更新技術や [18] のようなサーバ連携技術を，また，更新に伴うリグレッション試験には，例えば [19] のような自動検証技術を，用いることを

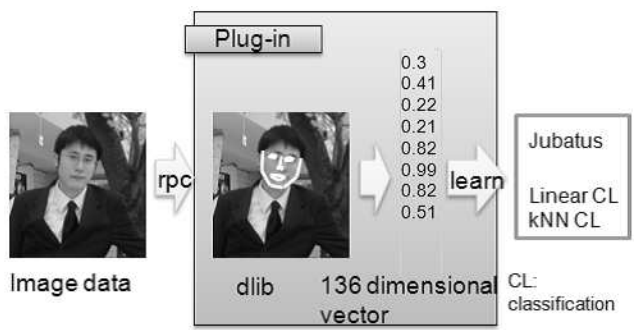

図 2  Jubatus による画像分析の簡易検証

想定している．

## 5. ま と め

本稿では，クラウド技術，データ分析技術を用いて，小規模店舗でも導入可能な，低コストな万引防止サービスを提案した．提案技術は，店舗側に配置された小型コンピュータ上の Jubatus が，監視カメラ映像を分析し異常を検出してクラウドに通知し，クラウド上の万引防止アプリケーションが，ERP 等の商品数データと突合し，万引の疑いが高い場合に，店舗スタッフにメール等で通知する．本稿では，まず，監視カメラ映像からのリアルタイムでの姿勢等の分析が Jubatus で可能かの簡易評価を行い，一定の正解率での判定ができること，しかし監視カメラの分析だけでは精度が不十分であり ERP 等との突合が必要であることを確認した．

今後は，万引防止の有効性を確認するため，実際のスーパー，ドラッグストア，書店等の店舗に提案し，トライアルを実施し検証する予定である．